\documentclass[twocolumn,showpacs,preprintnumbers,amsmath,amssymb,superscriptaddress]{revtex4-1}

\usepackage{graphicx}
\usepackage{dcolumn}
\usepackage{bm}

\begin{document}


\title{Influence of the liquid film thickness on the coefficient of restitution for wet particles}


\author{Thomas M\"uller}
\author{Kai Huang}
\email{kai.huang@uni-bayreuth.de}
\affiliation{Experimentalphysik V, Universit\"at Bayreuth, 95440 Bayreuth, Germany}

\date{\today}

\begin{abstract}

The normal coefficient of restitution (COR) for a spherical particle bouncing on a wet plane is investigated experimentally and compared with a model characterizing the energy loss at impact. For fixed ratios of liquid film thickness $\delta$ to particle diameter $D$, the wet COR is always found to decay linearly with St$^{-1}$, where St, the Stokes number, measures the particle inertia with respect to the viscous force of the liquid. Such a dependency suggests a convenient way of predicting the wet COR with two fit parameters: A critical COR at infinitely large St and a critical St at zero COR. We characterize the dependency of the two parameters on $\delta/D$ and compare it with a model considering the energy loss from the inertia and the viscosity of the wetting liquid. This investigation suggests an analytical prediction of the COR for wet particles.
\end{abstract}

\pacs{45.70.-n, 45.50.Tn, 47.55.Kf}

\maketitle

\section{Introduction}
\label{intro}

As large agglomerations of macroscopic particles, granular materials are ubiquitous in nature, industries and our daily lives~\cite{Nagel96,Duran00}. Due to the energy dissipation through particle-particle interactions, continuous energy injection is necessary to keep a granular material in a stationary state which is typically far from thermodynamic equilibrium. Thus, an important key to understand the dynamics of granular materials is to analyze the balance between energy injection and dissipation. For binary impacts, the coefficient of restitution, which was introduced by Newton~\cite{Newton1687} as the ratio between the relative rebound and impact velocities, provides a convenient way of characterizing the energy dissipation in fluidized granular systems~\cite{Jenkins83,Melo1995,Bizon98,Goldhirsch03,bril04,Huang2006,Huang2006a,Goetzendorfer2006,Aranson2006,Huang2010}. Over centuries, continuous investigations have led to substantial progresses in understanding how the energy is dissipated (e.g., through viscoelastic or plastic deformations~\cite{Love1927,Tabor1948,Johnson85,Rami99,Stronge2004,Antonyuk2010}). Moreover, the adhesive interactions arising from the surface energy of the deformed particles have also been considered in predicting the COR~\cite{Thornton98, Bril07}, using the well established Johnson-Kendall-Roberts (JKR) model~\cite{Johnson1971,Barthel2008}. 

Here, we focus on the case of a liquid film covering the solid bodies under impact, in order to shed light on the collective behavior of wet granular matter at a particle level. Recent investigations have revealed that clustering~\cite{Ulrich2009, Huang2012}, phase transitions~\cite{Fingerle2008,Huang2009a,May2013,Huang2015} as well as pattern formation~\cite{Huang2011, Butzhammer2015} of wet granular matter are often related to the `microscopic' particle-particle interactions, among which the wetting liquid plays an important role. Because the presence of a liquid film as thin as a few nanometers can be sufficient to influence the rigidity of granular matter substantially~\cite{Hornbaker1997, Huang2009}, it is essential to consider such an influence in the omnipresent applications. For example, it is associated with the modeling of natural disasters such as debris flow and volcano eruptions~\cite{Iverson1997,Telling2013}, and the granulation process in chemical engineering and pharmaceutics industries~\cite{Iveson01, Iveson01b}. 

In the past decades, there has been a growing interest in understanding the energy dissipation associated with wet impacts in order to predict the wet COR~\cite{Rumpf62,Ennis91,Iveson01,Davis02,Antonyuk09,Mueller11}. In the low Reynolds number regime where the viscosity of the wetting liquid dominates, the Stokes number was found to be the relevant parameter determining the influence of the size and density of the particles, as well as the viscosity of the liquid on the COR of binary as well as three-body collisions~\cite{Davis02,Donahue10}. The Stokes number is defined as ${\rm St}=\rho_{\rm p}Dv_{\rm i}/9\eta$ with particle density $\rho_{\rm p}$, particle diameter $D$,  impact velocity $v_{\rm i}$, and the dynamic viscosity of the liquid $\eta$. In the case of relatively high Reynolds number where the inertia of the liquid cannot be ignored, a former investigation~\cite{Gollwitzer2012} revealed that the dimensionless liquid film thickness $\tilde{\delta}=\delta/D$ (film thickness over particle diameter) starts to play an additional role. For $\tilde{\delta}\approx 0.04$, the dependency of the wet COR on various particle and liquid properties was characterized with the Stokes number~\cite{Mueller2013,Sutkar2014}. Despite of this progress, it is still unclear how the dimensionless film thickness influences the wet COR quantitatively. In this work, we explore this influence through a systematic tuning of $\tilde{\delta}$ in the experiments and compare the results with a model considering the liquid mediated energy loss during the impact. 

\section{Experimental Setup}
\label{sec:setup}

\begin{figure}
\includegraphics[width = 0.4\textwidth]{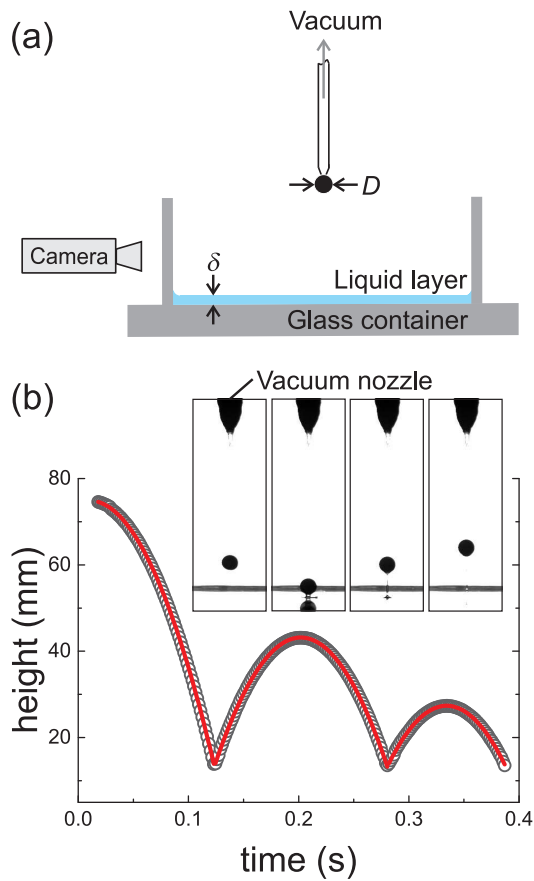}
\caption{\label{fig:setup}(color online) (a) Schematic of the free-fall experimental setup with a definition of the liquid film thickness $\delta$ and the particle diameter $D$. (b) The trajectory of a PTFE sphere ($D=8.000$\,mm) bouncing on a glass plate covered with a silicone oil film ($\delta=800\,\mu$m). Inset: Raw images taken (from left to right) before, during and after the first impact with a time step of $0.015$\,s. Solid curves in (b) correspond to parabolic fits to the individual bouncing events.}
\end{figure}

As illustrated in Fig.\,\ref{fig:setup}, we perform free-fall experiments to measure the normal COR of a wet spherical particle bouncing on the bottom of a rectangular glass container covered with a liquid film. The bottom plate is leveled within $0.03$\,degrees to ensure a homogeneous film thickness, which is measured optically from the deflection of an oblique laser beam shined from below the container. A more detailed description of this method can be found in Ref.~\cite{Mueller2013}. Two types of silicone oil (Wacker AK10 and Carl Roth M50, see Table\,\ref{tab:liq} for the specifications) are used as wetting liquids. Two types of particles (Spherotech, G2), polytetrafluorethylen (PTFE) with a density of $2.15$\,g$\cdot$cm$^{-3}$ and polyethylene (PE) with a density of $0.94$\,g$\cdot$cm$^{-3}$, are used. For each combination of particle type and wetting liquid, we vary systematically $\delta$ such that $\tilde{\delta}$ grows stepwise from $\sim0.03$ to $\sim0.15$. The diameters of the spherical particles are $D=3.969$, $4.762$ and $7.938$\,mm for PE particles, and $D=3.175$, $4.762$ and $8.000$\,mm for PTFE particles. The roughness of the particles is $\approx5$\,$\mu$m. The impact velocity is tuned via adjusting the initial falling height from $3$\,cm to $15$\,cm. 

After the wetting liquid is poured into the container, we wait for at least $30$ minutes for the liquid film thickness to become stable. The free fall motion of an initially wet particle is triggered by tuning the air pressure in the nozzle. When the free-falling particle enters the field of interest, a computer controlled high speed camera (Lumenera LT225) starts to take images. Subsequently, the images [see the inset of Fig.~\ref{fig:setup}(b) for an example] are subjected to an image analysis program that removes the background and detects the positions of a particle with sub-pixel resolution. As shown in Fig.\,\ref{fig:setup}(b), we fit each bouncing event with a parabola, from which the impact $v_{\rm i}$ as well as the rebound $v_{\rm r}$ velocities are determined. Based on its definition, the normal COR is obtained from $e_{\rm n}=v_{\rm r}/v_{\rm i}$. In order to have a well defined initial condition, only the COR from the first rebound is used in the analysis. For each falling height, at least five consecutive experimental runs are conducted with a waiting time of $\ge2$ minutes to ensure a stable $\delta$. More details on the experimental setup and procedure can be found in Ref.~\cite{Gollwitzer2012}. 

\begin{table}
\caption{\label{tab:liq}
Material properties of the wetting liquids at $25\,^{\circ}{\rm C}$.}
\begin{ruledtabular}
\begin{tabular}{cccc}
\textrm{}&
\textrm{Density}&
\textrm{Dynamic viscosity}\\
\textrm{}&
\textrm{(kg/m$^3$)}&
\textrm{(mPa\,s)}\\
\colrule
AK10 & 930 & 9.3\\
M50 & 965 & 48.3\\
\end{tabular}
\end{ruledtabular}
\end{table}

\section{Film thickness mediated scaling with Stokes number}
\label{sec:scal}

\begin{figure*}
\includegraphics[width = 0.45\textwidth]{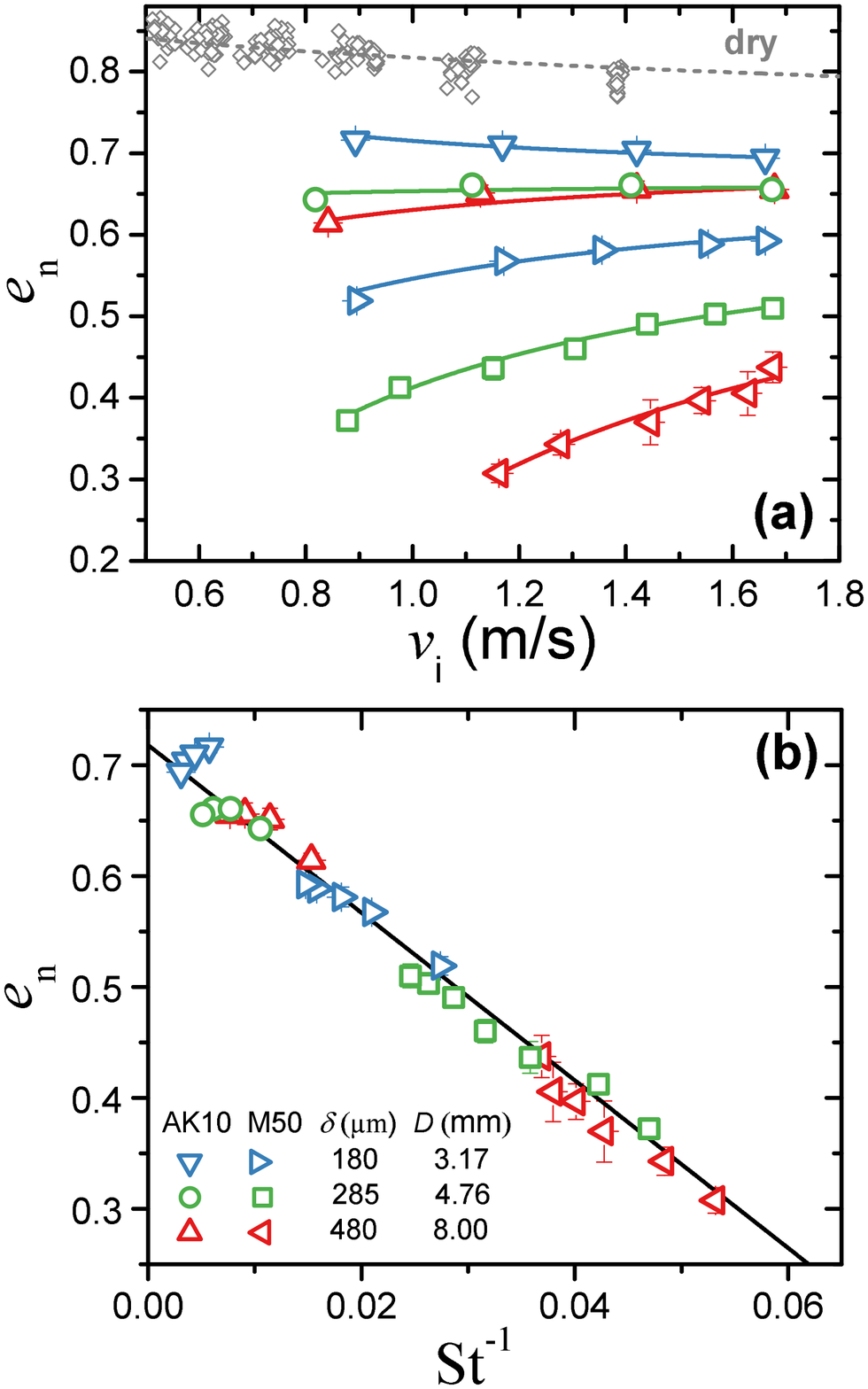}
\includegraphics[width = 0.45\textwidth]{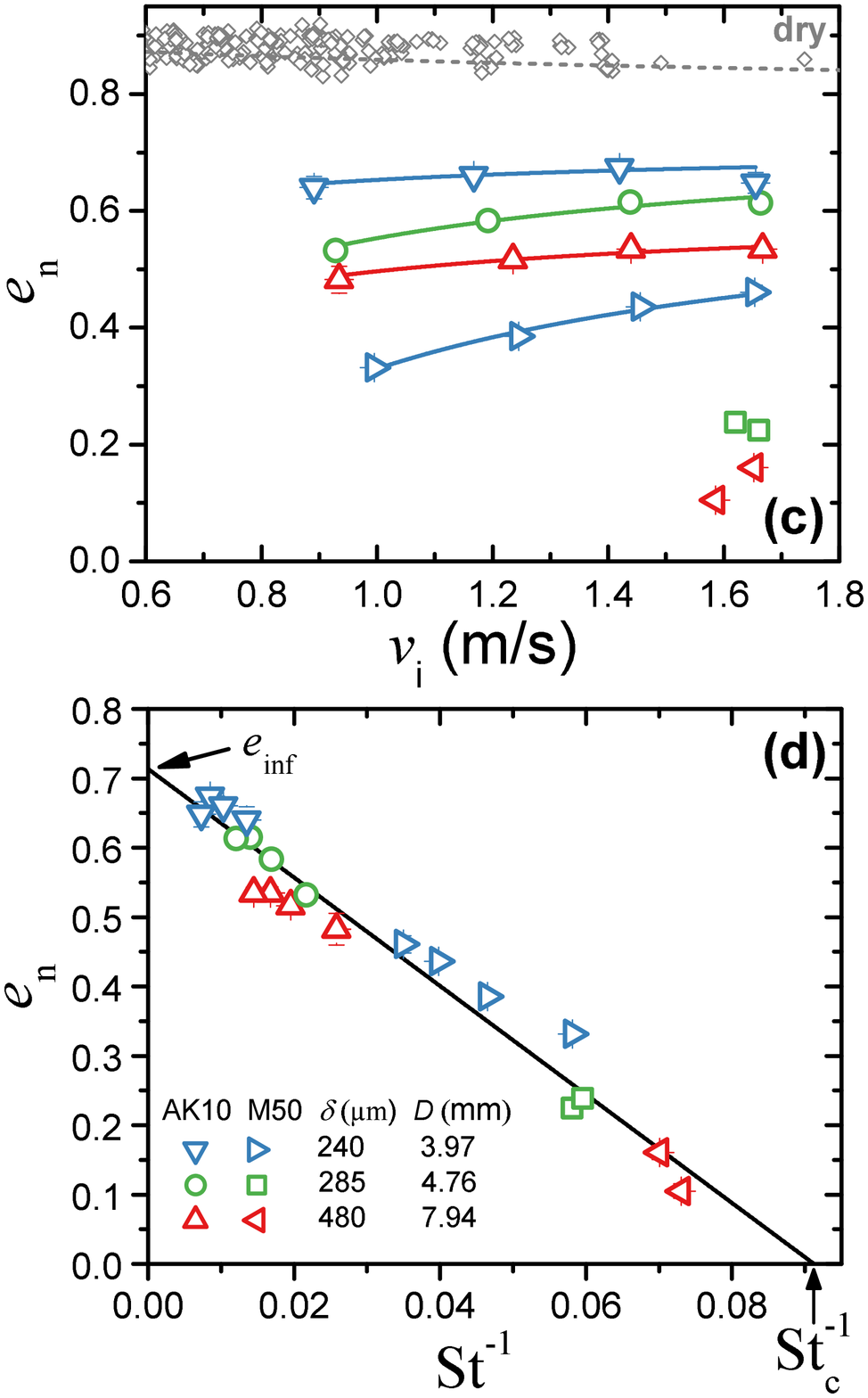}
\caption{\label{fig:envi}(color online) Upper panels: Normal coefficient of restitution $e_{\rm n}$ as a function of the impact velocity $v_{\rm i}$ measured with various wetting liquids (silicone oil AK10 and M50), $\delta$ and $D$ for both PTFE (a) and PE (c) particles. $\delta$ is chosen such that the dimensionless film thickness $\tilde{\delta}$ stays constant at $\approx0.06$. The dashed and solid lines are fits to the $e_{\rm n}$ obtained from the corresponding dry as well as wet impacts. Lower panels: $e_{\rm n}$ with respect to the inverted Stokes number St$^{-1}$ for PTFE (b) and PE (d) particles, respectively. A linear fit of the data obtained from all combinations of $\delta$ and $D$ (solid line) gives rise to two parameters: $e_{\rm inf}$ that represents the critical wet COR at ${\rm St}\to\infty$, and a critical Stokes number St${\rm_c}$ below which no rebound occurs. Fitting parameters are $e_{\rm inf}=0.718\pm 0.004$ and $0.713\pm 0.016$, ${\rm St}_{\rm c}=10.45\pm 0.53$ and $10.96\pm 0.54$ for PTFE and PE particles, respectively.}  
\end{figure*}

Before characterizing the wet COR and the associated energy dissipation from the wetting liquid, we measure the dry COR as a reference. As shown in the upper panels of Fig.\,\ref{fig:envi}, the dry COR decreases with the $v_{\rm i}$ for both PTFE and PE particles, in agreement with former experiments and theories~\cite{Bridges84,Montaine11,bril04,Antonyuk2010}. Qualitatively speaking, the maximum normal strain of the solid bodies reduces with growing impact velocity, therefore the smaller the $v_{\rm i}$, the closer the deformation is to an elastic one with $e_{\rm n}=1$. Following the nonlinear viscoelastic model~\cite{bril04} and taking the first order approximation, we fit the measured data with $1-kv_{\rm i}^{1/5}$ [dashed lines in Fig.\,\ref{fig:envi}(a) and (c)] and obtain $k=0.183\pm 0.001$ and $0.123\pm 0.001$ for PTFE and PE particles, respectively. The COR measured with larger $D$ yields a slightly smaller $e_{\rm n}$. However, the difference is small in comparison with the experimental uncertainty.

Figure~\ref{fig:envi}(a) shows the wet COR as a function of $v_{\rm i}$ for PTFE particles. For most cases, $e_{\rm n}$ grows monotonically with the impact velocity. The solid lines correspond to the fits of the data sets with $e_{\rm n} \propto v_{\rm i}^{-1}$. See the following Sec.~\ref{sec:theo} for a justification of the fit. For the case of $\delta=180\,\mu$m and less viscous AK10 wetting, $e_{\rm n}$ decays with the increase of $v_{\rm i}$. This exception is presumably due to the influence of the dry COR, because the energy loss from the wetting liquid decreases as the liquid film thickness or viscosity decreases. Indeed the decay shows the same trend as that of the dry COR (gray dashed curve), but with a shift of $\approx 0.1$. As the focus of this investigation is on the influence of the wetting liquid, we keep $\delta \ge 200\,\mu$m in the following analysis. The error bars, which represent the standard error arising from various runs of experiments, are within the size of the symbols for most of the parameters. Such a good reproducibility suggests that the initial condition of the particle (e.g., its degree of wetting) plays a minor role.

As shown in Fig.\,2(c), $e_{\rm n}$ obtained with PE particles and less viscous wetting liquid also grows monotonically with $v_{\rm i}$ and decreases with growing film thickness $\delta$. In agreement with the results obtained with PTFE particles, increasing the liquid viscosity or the film thickness yields smaller $e_{\rm n}$, since the energy dissipation through the viscous drag force increases. For the more viscous wetting liquid M50, less data points are obtained within the explored range of $v_{\rm i}$, because the particles hardly rebound, owing to the relatively small ratio of the particle inertia to the viscous force. 

In the lower panels of Fig.\,\ref{fig:envi}, we show the scaling of the wet COR with the Stokes number at a fixed $\tilde{\delta}=0.06$. For both types of particles, $e_{\rm n}$ obtained for various $\eta$, $D$, and $v_{\rm i}$ is found to decay linearly with ${\rm St}^{-1}$. Such a scaling reveals that the influences of liquid viscosity,  particle size, and impact velocity are coupled with each other through the Stokes number. The linear fit gives rise to two critical values: A critical wet COR $e_{\rm inf}$ at ${\rm St}\to\infty$ and a critical Stokes number ${\rm St}_{\rm c}$ below which no rebound occurs. Note that $e_{\rm inf}$ is smaller than $e_{\rm dry}$ for both PTFE and PE particles. Therefore we cannot estimate the saturated value of the wet COR at infinitely large $v_{\rm i}$ with $e_{\rm dry}$ if the liquid inertia does play a role (i.e., the Reynolds number is not sufficiently small). Here, the Reynolds number ${\rm Re}=\rho_{\rm l}\delta v_{\rm i}/\eta$ ranges from $6$ to $150$ at the beginning of impact. During the impact, ${\rm Re}$ decreases with $v_{\rm i}$, suggesting that the viscous drag force plays a more and more prominent role in comparison to the inertia of the liquid. Note that the Reynolds number and the Stokes number are coupled to each other with ${\rm Re}/{\rm St}=9\tilde{\rho}\tilde{\delta}$, where $\tilde{\rho} = \rho_{\rm l}/\rho_{\rm p}$ is the density ratio between the liquid and the particle.

\begin{figure}
\includegraphics[width = 0.45\textwidth]{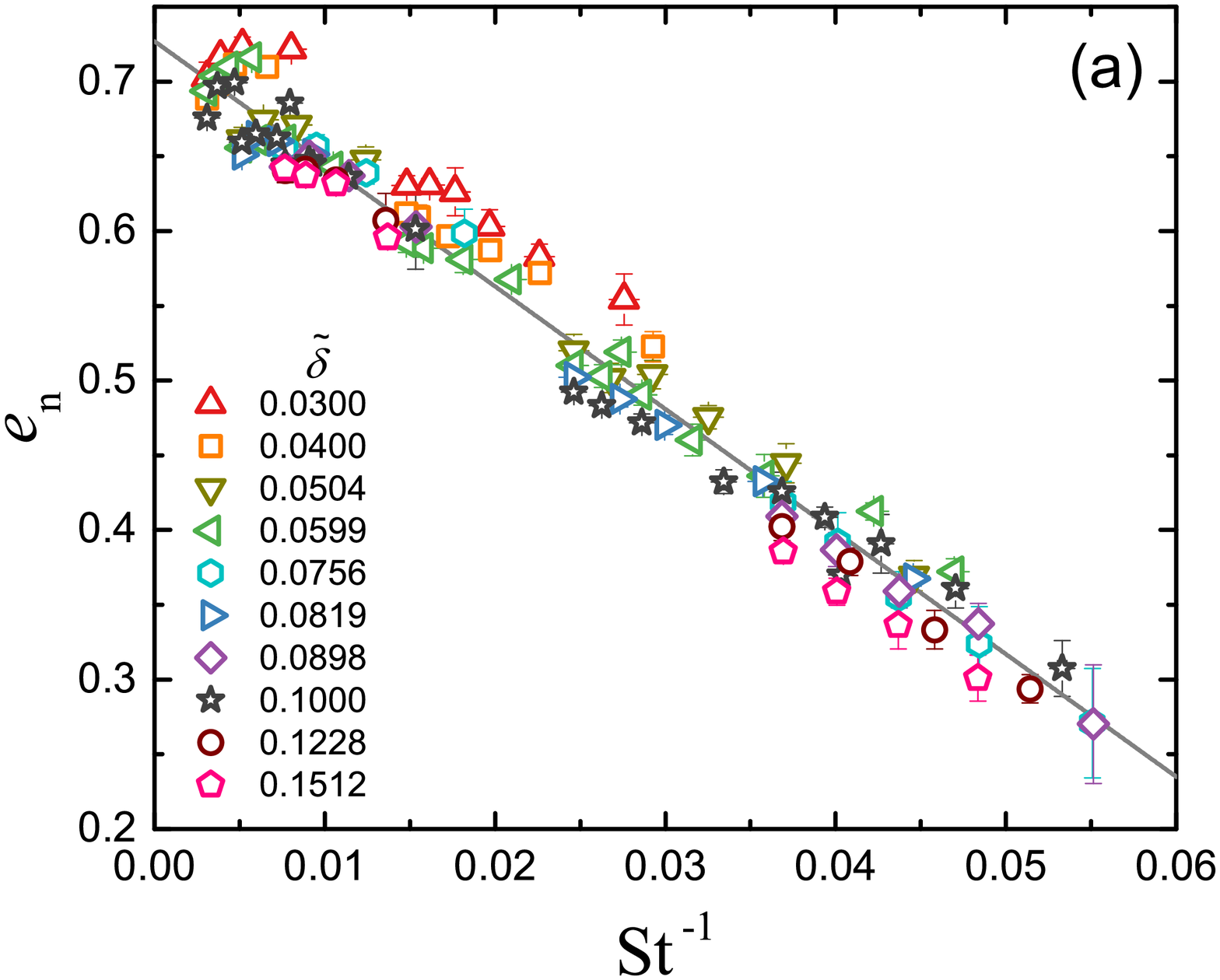} \\
\includegraphics[width = 0.45\textwidth]{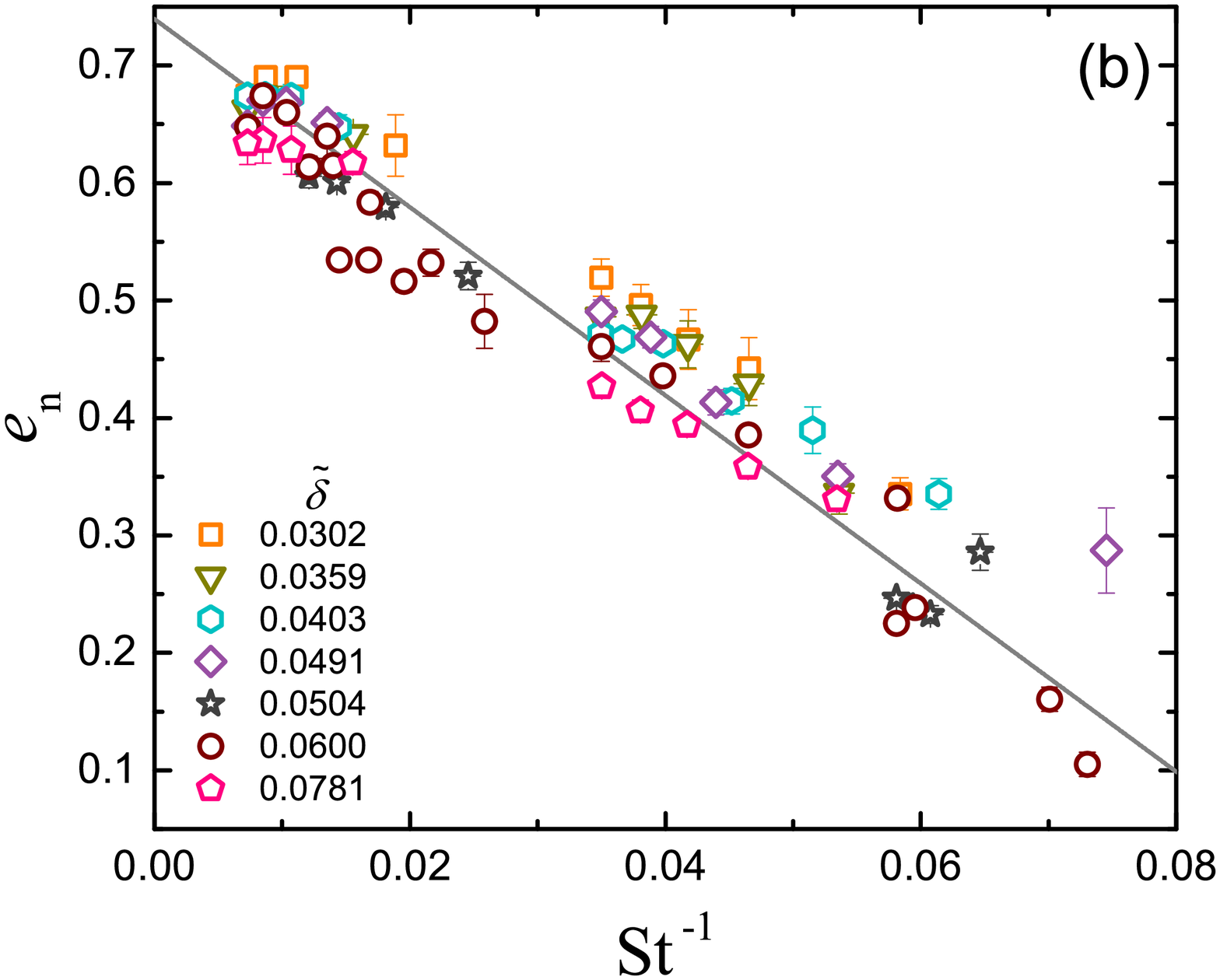}
\caption{\label{fig:varfilm}(color online) The wet COR $e_{\rm n}$ as a function of St$^{-1}$ for PTFE (a) and PE (b) particles at various $\tilde{\delta}$. $e_{\rm n}$ obtained with different particle and liquid properties are grouped according to the dimensionless film thickness $\tilde{\delta}$. The solid line corresponds to a linear fit to the data for all $\tilde{\delta}$. Fit parameters are: (a) $e_{\inf}=0.727\pm0.003$ and St$_{\rm c}=11.28\pm0.27$ for PTFE particles; (b) $e_{\inf}=0.740\pm0.006$ and St$_{\rm c}=10.83\pm0.26$ for PE particles. }
\end{figure}

Since the scaling of $e_{\rm n}$ with St suggests a convenient way of predicting the wet COR with $e_{\rm inf}$ and ${\rm St}_{\rm c}$, it is intuitive to step further and explore what determines the two fit parameters as well as possible ways to predict them. Motivated by this question, we vary systematically $\tilde{\delta}$ and check its influence on the scaling. 

Figure~\ref{fig:varfilm} shows the dependency of $e_{\rm n}$ with St$^{-1}$ for various dimensionless film thicknesses. For both PTFE (a) and PE (b) particles, the linear decay of $e_{\rm n}$ with St$^{-1}$ is prominent for all $\tilde{\delta}$. Moreover, the data obtained with various $\tilde{\delta}$ tend to collapse into a line. For PE particles, the upper limit of $\tilde{\delta}$ is smaller than that of PTFE particles, owing to the lack of rebound with thick liquid films. Note that in the lowest St$^{-1}$ region, $e_{\rm n}$ may grow with St$^{-1}$, particularly for the smallest $\tilde{\delta}$. This feature could be attributed to the influence from the dry COR, because, as we learned from the discussion of Fig.\,\ref{fig:envi}(a), the dependency of the dry COR on $v_{\rm i}$ dominates for a relatively thin and less viscous liquid film.

A closer analysis of the data reveals the influence of $\tilde{\delta}$: Data obtained with small $\tilde{\delta}$ lie above the fitted line, while data obtained with large $\tilde{\delta}$ do the opposite. In order to have a more quantitative analysis of such a dependency, we fit the data individually for each $\tilde{\delta}$. 

\begin{figure}
\includegraphics[width = 0.45\textwidth]{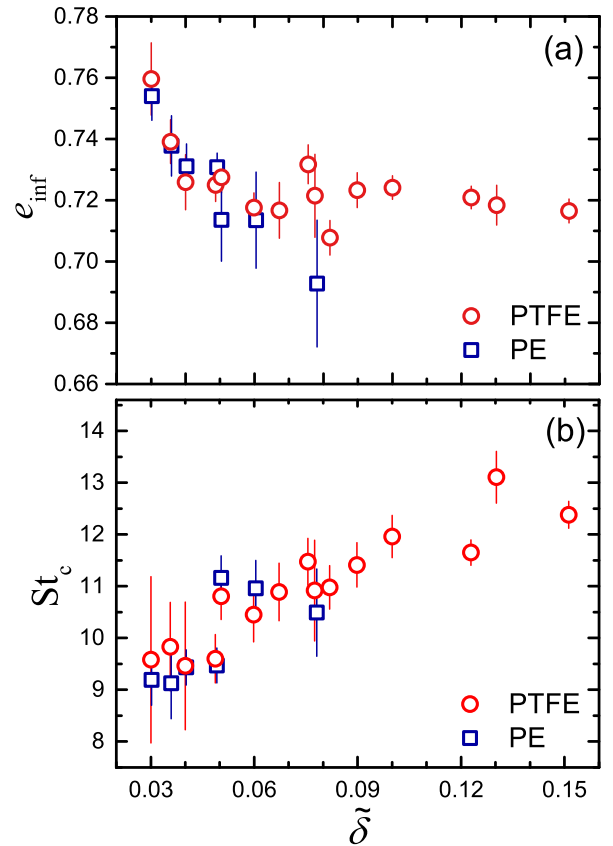}
\caption{\label{fig:par}(color online) Fit parameters $e_{\rm inf}$ (a) and ${\rm St}_{\rm c}$ (b) as a function of the dimensionless film thickness $\tilde{\delta}$ for both PTFE and PE particles.}
\end{figure}

Figure~\ref{fig:par} shows the fit parameters $e_{\rm inf}$ and ${\rm St}_{\rm c}$ as a function of $\tilde{\delta}$. As shown in (a), the critical wet COR decays monotonically with $\tilde{\delta}$ until it saturates at $e_{\rm inf}\approx 0.72$ for PTFE particles. For the case of PE particles, the range of $\tilde{\delta}$ is limited due to the reason described above. Within the limited range, the dependency of $e_{\rm inf}$ on $\tilde{\delta}$ agrees with the results obtained with PTFE particles within the error. As shown in (b), the critical Stokes number, ${\rm St}_{\rm c}$, grows monotonically with $\tilde{\delta}$ for both PTFE and PE particles. Within the common range of $\tilde{\delta}$, we also find a good agreement between the results obtained with PTFE and PE particles. Such an agreement suggests that the wetting liquid plays a dominant role in determining both fit parameters. At the limit of $\tilde{\delta}\to0$ (i.e., dry impact), we expect $e_{\rm inf}=e_{\rm dry}$ and ${\rm St}_{\rm c}=0$. As the dimensionless film thickness grows, the amount of energy taken by the inertia as well as viscosity of the liquid increases correspondingly. Therefore, we can qualitatively understand the trend of $e_{\rm inf}$ and ${\rm St}_{\rm c}$ as an indication of the enhanced energy loss from the liquid phase. In the following section, a more quantitative description of the influence will be presented.

\section{Model}
\label{sec:theo}

Following the above analysis, we present a model to explain the scaling with the Stokes number and discuss the possibility of predicting $e_{\rm inf}$ and ${\rm St}_{\rm c}$. 

According to its definition, the coefficient of restitution can be written as 

\begin{equation}
\label{eq:enEi}
e_{\rm n}=\sqrt{1-\frac{\Delta E_{\rm diss}}{E_{\rm i}}},
\end{equation}

\noindent where $E_{\rm i}=\frac{1}{2}\rho_{\rm p}V_{\rm p} v_{\rm i}^2$ with particle volume $V_{\rm p}$ is the kinetic energy of the particle before the impact and $\Delta E_{\rm diss}$ is the total amount of energy loss during the impact. $\Delta E_{\rm diss}$ includes the energy dissipation associated with inelastic solid body interactions~$\Delta E_{\rm dry}$ and the additional contribution from the wetting liquid $\Delta E_{\rm wet}$. Assuming that the two sources of kinetic energy loss are not coupled with each other, we have

\begin{equation}
\label{eq:enEdis}
e_{\rm n}=\sqrt{e_{\rm dry}^2-\frac{\Delta E_{\rm wet}}{E_{\rm i}}},
\end{equation}

\noindent where $e_{\rm dry}=\sqrt{1-\Delta E_{\rm dry}/E_{\rm i}}$ is the COR for dry impact. The energy loss from the wetting liquid $\Delta E_{\rm wet}$ has three main contributions: Surface energy due to the distorted liquid surface in both penetrating and rebouncing regimes, kinetic energy of the wetting liquid being mobilized $\Delta E_{\rm acc}$, and the energy dissipation from the viscous drag force $\Delta E_{\rm visc}$. Following a former investigation~\cite{Gollwitzer2012}, the rupture energy of a liquid bridge formed as a particle rebounds can be estimated with $\Delta E_{\rm b}\approx \pi \gamma \sqrt{2 V_{\rm b} D }$ with the liquid surface tension $\gamma$ and the volume of the liquid bridge $V_{\rm b}$. For the range of particle size explored here, this energy dissipation is ignorable because it is at least one order of magnitude smaller than $\Delta E_{\rm wet}$, even for the lowest $v_{\rm i}$~\cite{Gollwitzer2012}. Note that $\Delta E_{\rm b}$ is independent on $v_{\rm i}$, while both of the other two contributions grow with $v_{\rm i}$. Taking the other two terms into account, Eq.~\ref{eq:enEdis} can be rewritten as

\begin{equation}
\label{eq:enEdis2}
e_{\rm n} \approx \sqrt{e_{\rm dry}^2-\frac{\Delta E_{\rm acc}}{E_{\rm i}}-\frac{\Delta E_{\rm visc}}{E_{\rm i}}}.
\end{equation}

\noindent As the velocity of the liquid being pushed sidewards $v_{\rm l}$ arises from the penetration of the particle into the liquid film, we consider $v_{\rm l}\propto v_{\rm i}$ (see below for a more quantitative analysis). Consequently, we have $\Delta E_{\rm acc}\propto E_{\rm i}$. As the viscous drag force $\propto v_{\rm i}$, we consider the corresponding energy dissipation term $\Delta E_{\rm visc}\propto v_{\rm i}$. Thus, Eq.\,\ref{eq:enEdis2} can be rewritten as

\begin{equation}
\label{eq:envi}
e_{\rm n}=\sqrt{\alpha+\frac{\beta}{v_{\rm i}}}= \sqrt{\alpha}(1+\frac{\beta}{2\alpha}\frac{1}{v_{\rm i}}-\frac{\beta}{8\alpha}\frac{1}{v_{\rm i}^2}+\ldots).
\end{equation}

\noindent where $\alpha=e_{\rm dry}^2-\Delta E_{\rm acc}/E_{\rm i}$ and $\beta=-v_{\rm i}\Delta E_{\rm visc}/E_{\rm i}$ are $v_{\rm i}$ independent parameters. It indicates that the inertia of the wetting liquid contributes to a constant offset to $e_{\rm n}(v_{\rm i})$, while its combination with the viscous damping determines the factors of higher order terms. Since ${\rm St} \propto v_{\rm i}$, the linear decay of $e_{\rm n}$ with ${\rm St}^{-1}$ observed above can be treated as a first order approximation of Eq.\,\ref{eq:envi}. 

Moreover, Eq.\,\ref{eq:envi} can be used to predict the two fit parameters. On one hand, $e_{\rm inf}$, the wet COR at St$\to\infty$, can be estimated with

\begin{equation}
\label{eq:ec}
e_{\rm inf}=\sqrt{\alpha}=\sqrt{e_{\rm dry}^2-\frac{\Delta E_{\rm acc}}{E_{\rm i}}}.
\end{equation}

\noindent It shows that the saturated value of the wet COR is always smaller than $e_{\rm dry}$, in agreement with the experimental results shown in Fig.\,\ref{fig:envi}. Moreover, Eq.\,\ref{eq:ec} indicates that the difference between $e_{\rm dry}$ and $e_{\rm inf}$ arises from the energy taken by the inertia of the wetting liquid. 

On the other hand, a former analysis based on the lubrication theory~\cite{Gollwitzer2012} shows that

\begin{equation}
\label{eq:visc0}
\Delta E_{\rm visc}=\frac{3}{2}\pi \eta D^2 v_{\rm i} (\ln{\frac{\delta}{\epsilon}}+\ln{\frac{\delta_{\rm r}}{\epsilon}}),
\end{equation}

\noindent where $\delta_{\rm r}$ and $\epsilon$ are the rupture distance of the liquid bridge and the roughness of the particle, respectively. The two length scales arise from the limits of the separation distance within which the viscous force takes effect. In this estimation, it was assumed that the lubrication force applies during the whole impact period. This assumption becomes violated if the liquid film thickness is much larger than the critical separation distance $\delta_{\rm c}$, below which the lubrication theory applies~\cite{Davis02}. Since the lubrication theory predicts a diverging viscous force as the separation distance approaches $0$, we may consider that most of the energy loss due to the viscous force takes place within $\delta_{\rm c}$ and estimate the viscous damping term with

\begin{equation}
\label{eq:visc}
\Delta E_{\rm visc}=3\pi \eta D^2 v_{\rm i}\ln{\frac{\delta_{\rm c}}{\epsilon}}.
\end{equation}

\noindent Inserting it into the definition of $\beta$, we have

\begin{equation}
\label{eq:beta}
\beta=-\frac{36\eta v_{\rm i}}{\rho_{\rm p} D v_{\rm i}}\ln{\frac{\delta_{\rm c}}{\epsilon}}=-\frac{4v_{\rm i}}{\rm St}\ln{\frac{\delta_{\rm c}}{\epsilon}}.
\end{equation}

\noindent Note the essential role of the Stokes number here. Inserting Eqs.\,\ref{eq:ec} and \ref{eq:beta} into Eq.\,\ref{eq:envi} and taking the first order approximation, we have

\begin{equation}
\label{eq:en2}
e_{\rm n}=e_{\rm inf}(1-\frac{{\rm St_c}}{\rm St}),
\end{equation}

\noindent with the critical Stokes number 

\begin{equation}
\label{eq:stc}
{\rm St_c}=\frac{2\ln{\frac{\delta_{\rm c}}{\epsilon}}}{e_{\rm inf}^2}.
\end{equation}

\noindent Thus, the scaling of $e_{\rm n}$ with the Stokes number observed in the experiments is captured by the model. 

\begin{figure}
\includegraphics[width = 0.45\textwidth]{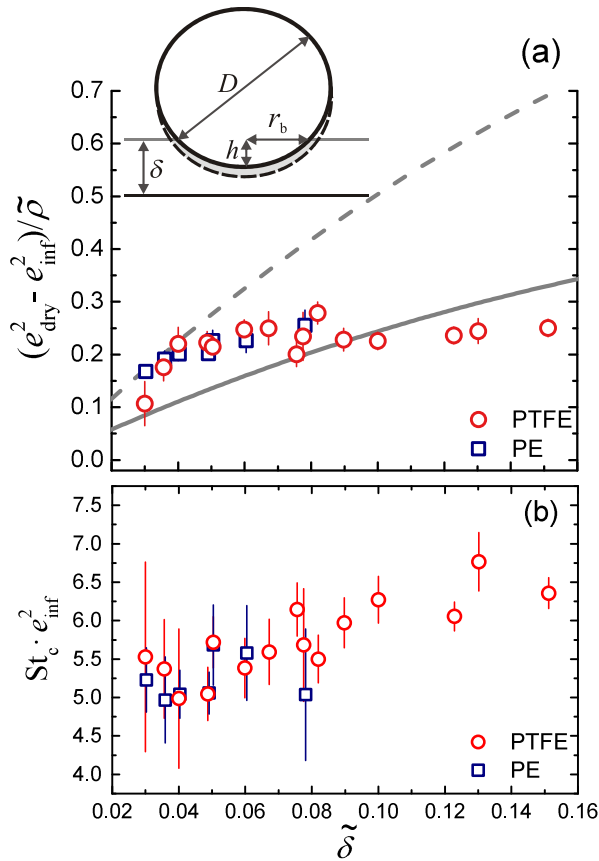} 
\caption{\label{fig:Eint}(color online) (a) The rescaled energy loss from the inertia of the liquid film as a function of the dimensionless film thickness $\tilde{\delta}$ for both types of particles. Different curves correspond to the predictions of different models describing the inertial effect: The dashed and solid curves correspond to the prediction of Eq.~\ref{eq:Eint1} and \ref{eq:Eint2}, respectively. The inset shows a sketch of a spherical particle penetrating into a liquid film. (b) Coupling between the critical Stokes number St$_{\rm c}$ and $e_{\rm inf}$ at various $\tilde{\delta}$, following the prediction of Eq.~\ref{eq:stc}.}
\end{figure}

As the next step, we discuss the dependency of the fit parameters on $\tilde{\delta}$. Starting from a former analysis~\cite{Gollwitzer2012}, we characterize the relative energy loss from the inertial effect with   

\begin{equation}
\label{eq:Eint1}
\frac{\Delta E_{\rm acc}}{E_{\rm i}}=\frac{2\rho_{\rm l}V_{\rm l}v_{\rm l}^2}{\rho_{\rm p}V_{\rm p}v_{\rm i}^2}\approx 2\tilde{\rho}\tilde{\delta}(3-5\tilde{\delta}+2\tilde{\delta}^2),
\end{equation}

\noindent where $V_{\rm l}$ is the volume of the liquid being expelled. The factor $2$ arises from the existence of inertial effects in both penetrating (liquid being repelled from the gap) and rebouncing (liquid being sucked into the gap) regimes. Here, the horizontal velocity of the liquid $v_{\rm l}$ is estimated with the base radius of the spherical cap over the penetrating time $\delta/v_{\rm i}$, where the particle is assumed to penetrate through the liquid film with the impact velocity $v_{\rm i}$. As sketched in the inset of Fig.\,\ref{fig:Eint}\,(a), the air liquid interface is assumed to be flat for the sake of simplicity. The additional influence from surface waves or the meniscus of a liquid bridge, which can lead to a modification of the kinetic energy being transferred from the particle to the liquid, will be a focus of further investigations.

Stepping further, we propose a more detailed model for $v_{\rm l}$, considering the stepwise approaching and receding of the particle. As illustrated in the inset of Fig.\,\ref{fig:Eint}(a), we consider the case of a spherical particle penetrating into a liquid film from a depth of $h$ (solid circle) to $h+dh$ (long dashed circle). Assuming the immersed part to be a spherical cap, we can estimate the volume of liquid being pushed sidewards with $V_{\rm cap}=\pi h^2(D/2-h/3)$ and the radius of the three phase contact line with the base radius $r_{\rm b}=\sqrt{h(D-h)}$. As the dimension of the container is much larger than that of the particle, we consider the film thickness $\delta$ to be constant during the impact. Consequently, we have $dV_{\rm cap}=\pi h(D-h)dh$ and a corresponding horizontal movement of

\begin{equation}
\label{eq:dl}
dr_{\rm b}=\frac{D-2h}{2\sqrt{h(D-h)}}dh.
\end{equation}

\noindent Due to momentum transfer, the liquid surrounding the particle is accelerated in the direction normal to the contact surface. However, the presence of the horizontal plane effectively guides the streamline to the horizontal direction. Suppose the change of flow direction is extremely efficient, we can estimate the velocity of the liquid being pushed sidewards with 

\begin{equation}
\label{eq:dl2}
v_{\rm l}(\tilde{h})=\frac{dr_{\rm b}}{dt}=\frac{0.5-\tilde{h}}{\sqrt{\tilde{h}(1-\tilde{h})}} v_{\rm i}(\tilde{h}),
\end{equation}

\noindent where $\tilde{h}=h/D$ is the dimensionless penetration depth. As $v_{\rm l}\propto v_{\rm i}$, the relative energy loss due to inertia of the liquid at each penetration step is independent of $v_{\rm i}$. Thus, an integration of $dE_{\rm acc}=\rho_{\rm l}v_{\rm l}^2dV_{\rm l}/2$ over the whole traveling distance leads to

\begin{equation}
\label{eq:Eint2}
\frac{\Delta E_{\rm acc}}{E_{\rm i}}=\frac{2\int dE_{\rm acc}}{E_{\rm i}}=\tilde{\rho} \tilde{\delta} (3-6\tilde{\delta}+4\tilde{\delta}^2).
\end{equation}

Note that in the receding regime, the flow of the liquid is reverted. Again, the factor $2$ arises from the assumption that the kinetic energy gained by the liquid in both approaching and receding regimes is the same.

According to Eq.~\ref{eq:ec}, $\Delta E_{\rm acc}/E_{\rm i}$ can be obtained experimentally with $e_{\rm dry}^2-e_{\rm inf}^2$. As Eq.~\ref{eq:Eint1} and \ref{eq:Eint2} both suggest that $\tilde{\rho}$ contributes only a constant factor in $\Delta E_{\rm acc}$, it is intuitive to compare the wet COR obtained with different types of particles using $(e_{\rm dry}^2-e_{\rm inf}^2)/\tilde{\rho}$. Here, we obtain $e_{\rm dry}$ from the fits of dry COR shown in Fig.~\ref{fig:envi}. Instead of the unrealistic value of $e_{\rm dry}=0$ at $v_{\rm i}\to\infty$, we choose the one at $v_{\rm i}=2$\,m/s, which corresponds to the upper limit of the impact velocity used in the dry COR measurements. 

As shown in Fig.\,\ref{fig:Eint}(a), such a comparison reveals a similar trend for both types of particles: A monotonic growth with $\tilde{\delta}$ followed by a saturated value of $\approx 0.25$. The results from both PTFE and PE particles agree with each other within the error. Such an agreement supports the outcome of the above analysis, i.e., the energy loss due to the inertia of the wetting liquid accounts for the difference between $e_{\rm dry}$ and $e_{\rm inf}$. As $\tilde{\rho}$ for PTFE and PE particles differs by a factor of $\sim 2.3$, the agreement also supports the scaling of the relative energy dissipation $\Delta E_{\rm acc}/E_{\rm i}$ with $\tilde{\rho}$. Moreover, a comparison with the predictions of the two models reveals that the simplified model originally introduced in Ref.~\cite{Gollwitzer2012} overestimates the influence from inertia, particularly for $\tilde{\delta}\ge 0.04$. The new model considering stepwise penetrations shown in Eq.~\ref{eq:Eint2} provides a better approximation, but it still cannot capture the saturation of $e_{\rm inf}$ at larger $\tilde{\delta}$. This is presumably due to the assumption that all the momentum transfer to the liquid ends up in the horizontal direction. In the future, more detailed investigations on the flow field inside the liquid film at impact are necessary to have a better prediction of $e_{\rm inf}$.    

Concerning the critical Stokes number, Eq.\,\ref{eq:stc} suggests that it depends on $\tilde{\delta}$ through its inverse proportionality with $e_{\rm inf}^2$, as well as on the ratio $\ln({\delta_{\rm c}/\epsilon})$. Because of the logarithmic scale, the latter influence is relatively weak. Therefore, one could consider St$_{\rm c}\propto e_{\rm inf}^{-2}$. As shown in Fig.~\ref{fig:Eint}(b), this argument is supported by the experimental results, because ${\rm St}_{\rm c}\cdot e_{\rm inf}^2$ stays roughly constant at $\approx 5.5$ for the common range of $\tilde{\delta}$ explored for both types of particles. Following Eq.~\ref{eq:stc}, this value corresponds to a critical separation distance of $\delta_{\rm c}\sim 100\,\mu$m. It is a reasonable value because, for all $\delta$ used in the experiments, the length scale associated with the wet region of the particle (i.e., base radius of the spherical cap immersed in the liquid $r_{\rm b}$) is at least one order of magnitude larger than $\delta_{\rm c}$. In the range of $\tilde{\delta}\ge 0.10$, ${\rm St}_{\rm c}\cdot e_{\rm inf}^2$ obtained with PTFE particles tends to grow slightly with $\tilde{\delta}$. This can be attributed to the dependency of $\delta_{\rm c}$ on the film thickness~\cite{Davis88}.

Finally, the above analysis leads to an analytical prediction of the wet coefficient of restitution as a function of ${\rm St}$:

\begin{equation}
\label{eq:wetcor}
e_{\rm n}=e_{\rm inf}-\frac{k}{e_{\rm inf}}\cdot \frac{1} {\rm St},
\end{equation}

\noindent where $k=2\ln\delta_{\rm c}/\epsilon$ can be treated as a constant factor for $\tilde{\delta}<0.10$, and $e_{\rm inf}$ can be estimated with

\begin{equation}
\label{eq:einf}
e_{\rm inf}=\sqrt{e_{\rm dry}^2-\tilde{\rho}\tilde{\delta}(3-6\tilde{\delta}+4\tilde{\delta}^2)}.
\end{equation}

Such a prediction will be helpful in large scale computer simulations of wet granular flow, and hence shed light on the widespread applications such as granulation process in chemical engineering, debris flow or volcano eruption in geophysics as well as multiphase flow in civil engineering~\cite{Telling2013,Antonyuk2013,Armanini2013}. 

\section{Conclusions}

To summarize, this investigation shows that the linear dependency of the COR for wet particle impacts with St$^{-1}$ is robust against a variation of the dimensionless liquid film thickness $\tilde{\delta}$, and such a dependency can be rationalized with a model considering the kinetic energy loss from the inertia as well as viscous force of the liquid. It suggests the possibility of predicting the wet COR with two fit parameters: the critical wet COR $e_{\rm inf}$ as St$\to\infty$ and the critical Stokes number ${\rm St}_{\rm c}$ for a rebound to occur. Based on a systematic variation of both film thickness and particle size, we discuss how $\tilde{\delta}$ influences the fit parameters. We find that $e_{\rm inf}$ is predominately determined by the inertia of the liquid. Considering the stepwise kinetic energy gain of the wetting liquid at impact, we present an analytical estimation of $e_{\rm inf}$. Moreover, the model predicts St$_{\rm c}\propto e_{\rm inf}^{-2}$ with a factor related to the ratio between two length scales; i.e., the  critical separation distance for the lubrication theory to apply and the roughness of the particle. Therefore, St$_{\rm c}$ can also be predicted analytically.

In the future, a more detailed analysis of the flow field as well as surface waves caused by the impact is necessary to clarify the discrepancy between the experiments and the model in order to have a more accurate determination of the wet coefficient of restitution. In addition, the influence from the cavitation dynamics~\cite{Marston2010b} should also be addressed. 

\begin{acknowledgments}
We thank Ingo Rehberg, Christof A. Kr\"ulle, Manuel Baur, and Simeon V\"olkel for inspiring discussions and a critical reading of the manuscript. This work is supported by the German Research Foundation through Grant No.~HU1939/2-1. 
\end{acknowledgments}

%

\end{document}